\documentclass[prd,aps,preprint,epsfig,floats]{revtex4}
\usepackage{graphics}
\usepackage{epsfig}

\usepackage{graphicx}
\begin{document}
\title{ Observable Neutron-Anti-Neutron Oscillation, Baryogenesis and High 
Scale Seesaw}

\author{\bf R. N. Mohapatra\footnote{Invited plenary talk presented at 
the WHEPP9 symposium in Bhubaneswar, India in January, 2006.} }
\affiliation{Department of Physics and Center for String and
Particle Theory, University of Maryland, College Park, MD 20742,
USA}
\date{May, 2006}
 \begin{abstract} Seesaw mechanism has been a dominant paradigm in the 
discussion of neutrino masses. I discuss how this idea can be tested via 
a baryon number violating process such as $N-\bar{N}$ oscillation. Since 
the expected seesaw scale is high and the $N-\bar{N}$ amplitude goes like 
$M^{-5}_{R}$, one might think that this 
process is not observable in realistic seesaw models for neutrino masses.
In this talk I show that in supersymmetric models, the above conclusion 
is circumvented leading to an enhanced and observable rate for $N-\bar{N}$
oscillation. I also discuss a new mechanism for baryogenesis in generic 
 models for neutron-anti-neutron oscillation and show how the requirement 
of adequate baryogenesis can put an upper limit on the 
neutron-anti-neutron oscillation time. \end{abstract}

\maketitle
\section{Introduction}
There are various reasons to suspect that baryon number is not a
good symmetry of nature: (i) first is that nonperturbative
effects of the standard model lead to $\Delta B\neq 0$, while
keeping $\Delta (B-L)=0$  \cite{thooft}; (ii) second has to do with an 
understanding of the
origin of matter in the universe requires $\Delta B\neq 0$
interactions\cite{sakharov} and (iii) many theories beyond the standard 
model lead to interactions that violate baryon number  \cite{ps,gg}.

If indeed such interactions are there, an important question is:
can we observe them in experiments$\,$? Two interesting baryon
nonconserving processes of experimental interest are: (a) proton
decay e.g. $p\rightarrow e^++\pi^0, \bar{\nu}+K^0$ etc
\cite{gg,pdecay} and (b) $N\leftrightarrow \bar{N}$ oscillation
\cite{kuzmin,glashow,marshak}. These two classes of processes
probe two different selection rules for baryon nonconservation:
$\Delta (B-L)=0$ for proton decay and $\Delta(B-L)=2$ for
$N\leftrightarrow \bar{N}$ oscillation. They are signatures of two 
totally different directions for unification beyond the standard model.
For example, observation of
 proton decay will point strongly towards a grand desert till
 about the scale of $10^{16}$ GeV whereas $N\leftrightarrow \bar{N}$
oscillation will require new physics at an intermediate scale at or above the
 TeV scale but much below the GUT scale. Further experimental search for 
both these processes can therefore provide key insight into the nature of 
unification beyond the standard model with or without supersymmetry.  

 While proton decay goes very naturally with the idea of eventual
 grand unification of forces and matter, recent discoveries of neutrino
  oscillations have made $N\leftrightarrow \bar{N}$ oscillation to be
quite plausible theoretically if small neutrino masses are to be
understood as a consequence of the seesaw mechanism~\cite{seesaw}.
This can be seen as follows: seesaw mechanism implies Majorana
neutrinos implying the existence of $\Delta (B- L)=2$
interactions. In the domain of baryons, it implies the existence
of $N\leftrightarrow \bar{N}$ oscillation as noted many years ago
\cite{marshak}.
In fact an explicit model for $N\leftrightarrow \bar{N}$
oscillation was constructed in  \cite{marshak} by implementing
the seesaw mechanism within the framework of the Pati-Salam
\cite{ps} $SU(2)_L\times SU(2)_R\times SU(4)_c$ model, where
quarks and leptons are unified. It was shown that this process is
mediated by the exchange of diquark Higgs bosons giving an
amplitude (see Fig. 1)
 ($G_{N \leftrightarrow \bar N}$) which scales like $M^{-5}_{qq}$. In the
nonsupersymmetric version without fine tuning, one expects
$M_{qq}\propto v_{BL}$ leading to $G_{N \leftrightarrow \bar
N}\simeq v^{-5}_{BL}$. So only if $M_{qq}\sim v_{BL}\sim 10 -100$
TeV, the
  $\tau_{N\leftrightarrow \bar{N}}$ is in the range of
$10^6-10^{8}$ sec and is accessible to experiments. On the other
hand, in generic seesaw models for neutrinos, one expects
$v_{BL}\sim 10^{11}-10^{14}$ GeV depending on the range of the
third generation Dirac mass for the neutrino of 1-100 GeV. An
important question therefore is whether in realistic seesaw
models, $N\leftrightarrow \bar{N}$ oscillation is at all
observable. Another objection to the above
nonsupersymmetric model for $ N \leftrightarrow \bar N$ that was
raised in the  80's was that such interactions will erase any
baryon asymmetry created at high scales. It is therefore
important to overcome this objection.

Several years ago, a high scale seesaw model with observable $N-\bar{N}$ 
oscillation
was presented using R-parity violating interactions\cite{babu}. Such 
models in general lead to difficulties in understanding the origin of 
matter and also do not have a naturally stable supersymmetric dark 
matter. 

 In this talk, I first discuss a recent paper\cite{DMM} where it is shown 
that in a class of supersymmetric $SU(2)_L\times
SU(2)_R\times SU(4)_c$ models (called SUSY $G_{224}$), an
interesting combination of circumstances improves the $v_{BL}$
dependence of the $G_{\Delta B=2}$ to $v^{-2}_{BL}v^3_{wk}$
instead of $v^{-5}_{BL}$ making $N\leftrightarrow \bar{N}$
oscillation observable. This does not require the existence of R-parity 
violation and in fact in these models R-parity is naturally conserved 
giving rise to a stable dark matter. I then discuss a new mechanism for 
post-sphaleron 
baryogenesis where an upper limit on $N-\bar{N}$ oscillation time is 
required to 
generate the desired baryon to photon ratio of the Universe.
 
The basic ingredients of such a theory was presented in  \cite{chacko}
 where it was shown that in the minimal supersymmetric $SU(2)_L\times
SU(2)_R\times SU(4)_c$ model, there exist accidental symmetries that 
imply that some of the $M_{qq}$'s which mediate $N-\bar{N}$ oscillation 
are in the TeV range even though $v_{BL}\simeq 10^{11}-10^{12}$ GeV. The 
present work\cite{DMM} points out that there exist a new class of Feynman 
diagrams which enhance the $N\leftrightarrow \bar{N}$ oscillation 
amplitude in generic supersymmetric models of this type making it 
observable. I then discuss the new mechanism for baryogenesis in 
models where the baryon number violation is mediated by a higher 
dimensional operator such as in the case of $N-\bar{N}$ 
oscillation\cite{babu1}.

At present, the best lower bound on $\tau_{N\leftrightarrow
\bar{N}}$ comes from ILL reactor experiment \cite{milla} and is
$10^{8}$ sec. There are also comparable bounds from nucleon decay
search experiments \cite{nnbar}. There are proposals to improve
the precision of this search by at least  two orders of magnitude
\cite{kamyshkov}. We feel that the results of this paper\cite{DMM,babu1} 
should give new impetus to a search for neutron-antineutron oscillation.

\section{$SU(2)_L\times SU(2)_R\times SU(4)_c$ model with light
diquarks}

The quarks and leptons in this model are unified and transform as
$\psi:({\bf 2,1,4})\oplus \psi^c:({\bf 1,2},\bar{\bf 4})$
representations of $SU(2)_L\times SU(2)_R\times SU(4)_c$. For the
Higgs sector, we choose, $\phi_1:(\bf{2,2,1})$ and
$\phi_{15}:(\bf{2,2,15})$ to give mass to the fermions. The
$\Delta^c:({\bf 1,3,10})\oplus \bar{\Delta}^c:({\bf
1,3},\overline{\bf 10})$ to break the $B-L$ symmetry. The diquarks
mentioned above which lead to $\Delta (B-L)=2$ processes are
contained in the $\Delta^c:(\bf{1,3,10})$ multiplet. We also add a
$B-L$ neutral triplet $\Omega:(\bf{1,3,1})$ which helps to reduce
the number of light diquark states.
 The superpotential of this model is given
by:
\begin{eqnarray}
W~=~W_Y~+~W_{H1}+~W_{H2}+~W_{H3}
\end{eqnarray}
where
\begin{eqnarray}
W_{H1}&=&\lambda_1 S( \Delta^c\bar{\Delta}^c-M^2)~+\lambda_C 
\Delta^c\bar{\Delta}^c\Omega~ +\mu_{i}{\rm Tr}\,(\phi_i\phi_i)
\end{eqnarray}
\begin{eqnarray}
W_{H2}~=~\lambda_A
\frac{(\Delta^c\bar{\Delta}^c)^2}{M_{P\!\ell}}
+ \lambda_B\frac{(\Delta^c{\Delta^c}) 
(\bar{\Delta}^c\bar{\Delta}^c)}{M_{P\!\ell}}
\end{eqnarray}
\begin{eqnarray}
 W_{H3}~=~ \lambda_D \frac{{\rm Tr}\,(\phi_1\Delta^c
\bar{\Delta}^c\phi_{15})}{M_{P\!\ell}} \,, \\
\end{eqnarray}
\begin{eqnarray}
W_Y&=&h_1\psi\phi_1 \psi^c + h_{15} \psi\phi_{15} \psi^c + f
\psi^c\Delta^c \psi^c.
\end{eqnarray}
Note that since we do not have parity symmetry in the model, the
Yukawa couplings $h_1$ and $h_{15}$ are not symmetric matrices.
When $\lambda_B=0$, this superpotential has an accidental global
symmetry much larger than the gauge group\cite{chacko}; as a
result, vacuum breaking of the $B-L$ symmetry leads to the
existence of light diquark states that mediate $n\leftrightarrow
\bar{n}$ oscillation and enhance the amplitude. In fact it was
shown that for $\langle\Delta^c\rangle\sim
\langle\bar{\Delta}^c\rangle\neq 0$ and $\langle\Omega\rangle\neq
0$ and all VEVs in the range of $10^{11}-10^{12}$ GeV, the light
states are those with quantum numbers: $\Delta_{u^cu^c}$.
 The symmetry argument behind is that \cite{chacko}
 for $\lambda_B=0$, the above superpotential is invariant under
$U(10,c)\times SU(2,c)$ symmetry  which breaks down to
$U(9,c)\times U(1)$ when $\langle\Delta^c_{\nu^c\nu^c}\rangle =
v_{BL}\neq 0$. This results in 21 complex massless states; on the
other hand these vevs also breaks the gauge symmetry down from
$SU(2)_R\times SU(4)_c$ to $SU(3)_c\times U(1)_Y$. This allows
nine of the above states to pick up masses of order $gv_{BL}$
leaving 12 massless complex states which are the six
$\Delta^c_{u^cu^c}$ plus six $\bar{\Delta}^c_{u^cu^c}$ states.
Once $\lambda_B\neq 0$ and is of order $10^{-2}-10^{-3}$, they
pick up mass (call $M_{u^cu^c}$) of order of the elctroweak scale.

\section{$N\leftrightarrow \bar{N}$ oscillation- a new diagram}

To discuss $N\leftrightarrow \bar{N}$ oscillation,
we introduce a new term in the superpotential of the form\cite{marshak}:
\begin{eqnarray}
W_{\Delta B=2}~=~\frac{1}{M_*}\epsilon^{\mu'\nu'\lambda'\sigma'}
\epsilon^{\mu\nu\lambda\sigma}
\Delta^c_{\mu\mu'}\Delta^c_{\nu\nu'}\Delta^c_{\lambda\lambda'}
\Delta^c_{\sigma\sigma'}\,, \label{Delta_B=2}
\end{eqnarray}
where the $\mu,\nu  $ etc stand for $SU(4)_c$ indices and we have
suppressed the $SU(2)_R$ indices. Apriori $M_*$ could be of order
$M_{P\ell}$; however the terms in Eq.(2) are different from those
in Eq. (4); so they could arise from different a high scale
theory. The mass $M_*$ is therefore a free parameter that we
choose to be much less than the $M_{P\ell}$. This term does not
affect the masses of the Higgs fields. When
$\Delta_{\nu^c\nu^c}^c$ acquires a VEV, $\Delta B = 2$ interaction
are induced from this superpotential,
%
and $N\leftrightarrow \bar{N}$ oscillation are generated by two
diagrams given in Fig. 1 and 2. The first diagram (Fig. 1) in
which only diquark Higgs fields are involved was already discussed
in  \cite{marshak} and goes like $G_{N\leftrightarrow \bar
N}\simeq \frac{f^3_{11}v_{BL}
M_\Delta}{M^2_{u^cu^c}M^4_{d^cd^c}M_*}$,
 Taking $M_{u^cu^c}\sim 350$ GeV, $M_{d^cd^c}\sim \lambda'v_{BL}$ and
$M_\Delta \sim v_{BL}$ as in the argument \cite{chacko}, we see
that this
diagram scales like $v^{-3}_{BL} v_{wk}^{-2}$. 

In ref.\cite{DMM} a new diagram (Fig. 2) was pointed out which owes its 
origin to supersymmetry. We get for its contribution to $G_{\Delta 
B=2}$:
\begin{eqnarray}
G_{N \leftrightarrow \bar N}\simeq\frac{g^2_3}{16\pi^2}
\frac{f^3_{11}v_{BL}}{M^2_{u^cu^c}M^2_{d^cd^c}M_{\rm SUSY}M_*}.
\end{eqnarray}
Using the same arguments as above, we find that this diagram
scales like $v^{-2}_{BL} v_{wk}^{-3}$ which is therefore a significant
enhancement over diagram in Fig.1.
\begin{figure}[h!]
\includegraphics[scale=0.8]{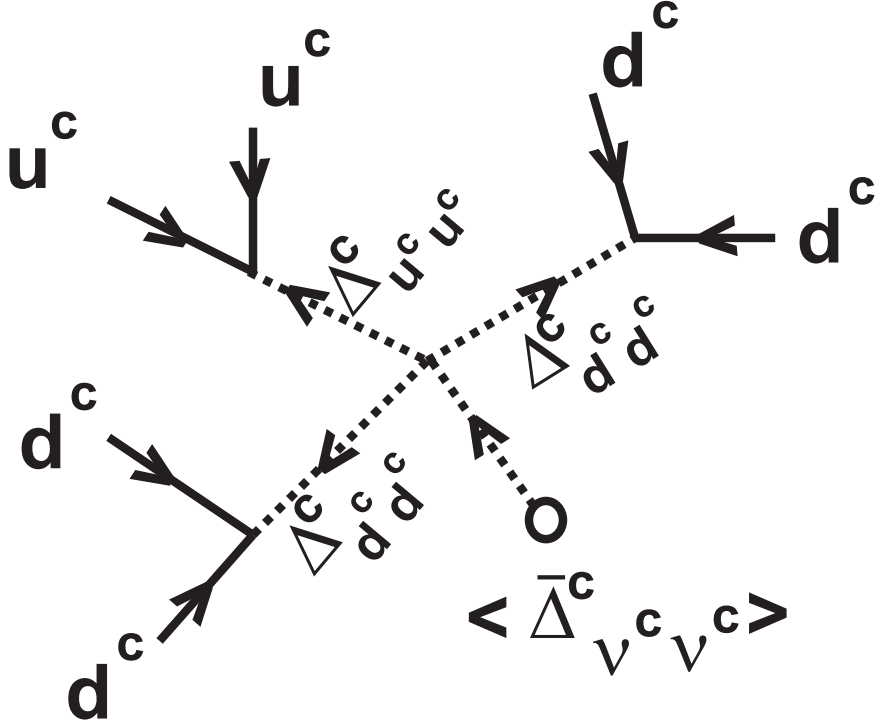}
\caption{The Feynman diagram responsible for $N-\bar{N}$ oscillation as
discussed in Ref. 8. }
\end{figure}

\begin{figure}[h!]
\includegraphics[scale=0.8]{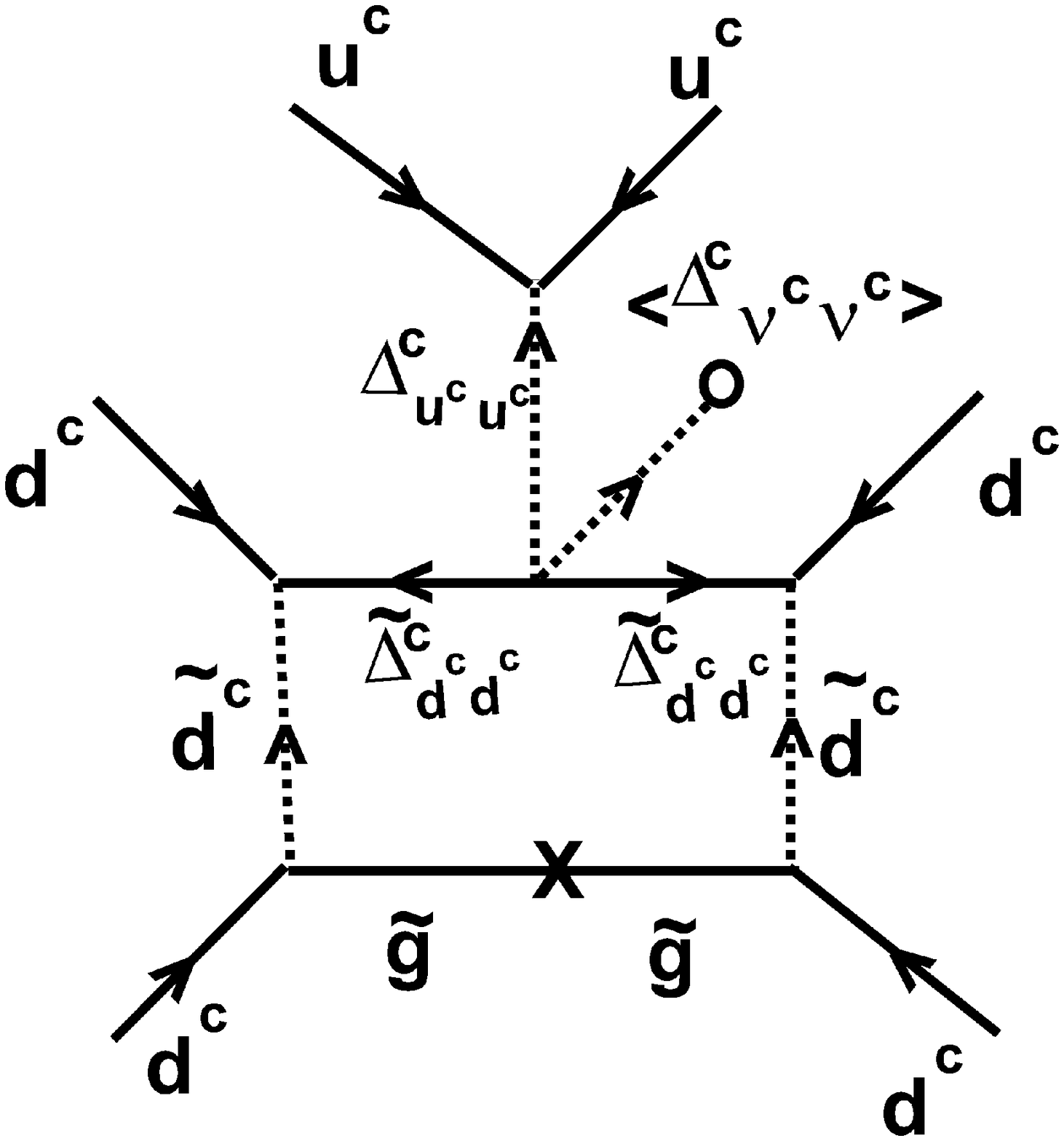}
\caption{The new Feynman diagram for $N-\bar{N}$ oscillation.}
\end{figure}
In order to estimate the rate for $N\leftrightarrow \bar{N}$
oscillation, we need not only the different mass values for which
we now have an order of magnitude, we also need the Yukawa
coupling $f_{11}$. Now $f_{11}$ is a small number since its value
is associated with the lightest right-handed neutrino mass.
However, in the calculation we need its value in the basis where
quark masses are diagonal. We note that the
 $N-\bar{N}$ diagrams involve only the
right-handed quarks, the rotation matrix need not be the CKM
matrix. The right-handed rotations need to be large e.g. in order
to involve $f_{33}$ (which is $O(1)$), we need
$(V_R^{(u,d)})_{31}$ to be large, where $V_L^{(u,d)\dagger}
Y_{u,d}V_R^{(u,d)}=Y_{u,d}^{\rm diag.}$. The left-handed rotation
matrices $V_L^{(u,d)}$ contribute to the CKM matrix, but
right-handed rotation matrices $V_R^{(u,d)}$ are unphysical in the
standard model. In this model, however, we get to see its
contribution since we have a left-right gauge symmetry.

Let us now estimate the time of oscillation. When we start with a
$f$-diagonal basis (call the diagonal matrix $\hat f$), the
Majorana coupling $f_{11}$ in the diagonal basis of up- and
down-type quark matrices can be written as $(V_R^T {\hat f}
V_R)_{11} \sim (V^R_{31})^2 \hat f_{33}$. Now $\hat f_{33}$ is
$O(1)$ and $V^R_{31}$ can be $\sim 0.6$, so $f_{11}$ is about 0.4
in the diagonal basis of the quark matrices. We use $M_{\rm
SUSY}$, $M_{u^cu^c}$ $\sim 350$ GeV and $v_{BL}\sim 10^{12}$ GeV.
 The mass of $\tilde\Delta_{d^cd^c}$ i.e. $M_{
d^cd^c}$ is $10^{9}$ GeV which is obtained from the VEV of
$\Omega:(\bf{1,3,1})$. We choose $M_*\sim 10^{13}$ GeV. Putting
all the above the numbers together, we get
\begin{eqnarray}
G_{N \leftrightarrow \bar N}\simeq 1 \cdot 10^{-30} \ {\rm
GeV^{-5}}. \label{G-NNbar}
\end{eqnarray}
Along with the hadronic matrix element \cite{Rao:1982gt}, the
$N-\bar{N}$ oscillation time is found to be about $2.5\times
10^{10}$ sec which is within the reach of possible next generation
measurements. If we chose, $M_*\simeq M_{P\ell}$, we will get for
$\tau_{N-\bar{N}}\sim 10^{15}$ sec. unless we choose the
$M_{d^dd^c}$ to be lower (say $10^7$ GeV). This is a considerable
enhancement over the nonsupersymmetric model of \cite{marshak}
with seesaw scale of $10^{12}$ GeV.

We also note that as noted in \cite{marshak} the model is
invariant under the hidden discrete symmetry under which a field
$X \rightarrow e^{i\pi B_{X}}X$, where $B_X$ is the baryon number
of the field $X$. As a result, proton is absolutely stable in the
model. Furthermore, since R-parity is an automatic symmetry of
MSSM, this model has a naturally stable dark matter.

\section{Baryogenesis and $N-\bar{N}$ oscillation}
In the early 1980's when the idea of neutron-anti-neutron
oscillation was first proposed in the context of unified gauge
theories, it was thought that the high dimensionality of the
$\Delta B\neq 0$ operator would pose major difficulty in
understanding the origin of matter. The main reason for this is
that the higher dimensional operators remain in thermal
equilibrium until late in the evolution of the universe since the 
thermal decoupling temperature $T_*$ for such interactions goes roughly 
like $v_{BL}\left(\frac{v_{BL}}{M_{P}}\right)^{1/9}$ which is in the 
range of temperatures where B+L violating sphaleron transitions are full 
thermal equilibrium. They will therefore erase any baryon 
asymmetry generated in the very early moments of the universe
(say close to the GUT time of $10^{-30}$ sec. or so) in prevalent 
baryogenesis models. In models
with observable $N-\bar{N}$ oscillation therefore, one has to
search for new mechanisms for generating baryons below the weak
scale. In this section, we discuss such a possibility\cite{babu1} 
discussed in a recent unpublished work with K. S. Babu and S. Nasri.

As an illustration of the way the new mechanism operates, let us assume 
that there is a scalar field that couples to the $\Delta B=2$ operator 
i.e. $L_I~=~Su^cd^cd^cu^cd^cd^c/M^6$, where the scalar boson has mass of 
order of the weak scale and $B=2$. This leads to baryon number
violation if $<S>\neq 0$ and observable $N-\bar{N}$ transition if $M$ is 
in the few hundred to 1000 GeV range. The direct decay of $S$ in these 
models can lead to an adequate mechanism for baryogenesis. 

To discuss how this comes about, let us first note that the high 
dimension of $L_I$
allows the scalar $\Delta B\neq 0$ decay to go out of equilibrium
at weak scale temperatures. This clearly satisfies the out of equilibrium 
condition given by Sakharov conditions for origin of matter\cite{sakharov}.
This is the reason we require a higher dimensional operator. For direct 
proton decay operators such as $QQQL$, the decoupling temperature is much 
higher and our mechanism will not apply.

To outline the rest of the details of this mechanism\cite{babu1}, we 
consider an effective sub-TeV scale model that gives rise to the 
higher dimensional
operator  for ${N\leftrightarrow \bar{N}}$ oscillation. It consists of
the following color sextet, $SU(2)_L$
singlet scalar bosons $(X,Y,Z)$ with hypercharge $-\frac{4}{3},
+\frac{8}{3}, +\frac{2}{3}$ respectively that couple to quarks.
These states could emerge from the supersymmetric model described in the 
previous section if $<\Omega>=0$.
We add to it a scalar field with mass in the 100 GeV range.
One can now write down the following standard model invariant
interaction Lagrangian:
\begin{eqnarray}
{\cal L}_I~&=&~ h_{ij}X d^c_id^c_j + f_{ij} Yu^c_iu^c_j~+
\\\nonumber && g_{ij} Z (u^c_id^c_j+u^c_jd^c_j)  +~\lambda_1 S
X^2Y~+~\lambda_2 SXZ^2
\end{eqnarray}
The scalar field $S$ is assumed to have $B=2$. To see the constraints on 
the parameters of the theory, we note
that the present limits on $\tau_{N-\bar{N}}\geq 10^8$ sec.
implies that the strength $G_{N-\bar{N}}$ of the the $\Delta B= 2$
transition is $\leq 10^{-28}$ GeV$^{-5}$. From Fig. 3, we conclude
that
\begin{eqnarray}
G_{N-\bar{N}}\simeq \frac{\lambda_1 M_1 h^2_{11}
f_{11}}{M^2_YM^4_X}~+~\frac{\lambda_2 M_1h_{11}g^2_{11}}{M^2_XM^2_Z}\leq
10^{-28} GeV^{-5}.
\end{eqnarray}
For $\lambda_{1,2} \sim h\sim f\sim g\sim 10^{-3}$, we have $M_{1}\sim
M_{X,Y,Z}\simeq 1$ TeV. In our discussion, we will stay close to
this range of parameters and see how one can understand the baryon
asymmetry of the universe. The singlet field will play a key role in the 
generation of baryon
asymmetry. We assume that $<S>\sim M_{X}$ but $M_{S_r}\sim 100$
GeV, where $S_r$ is the real part of the $S$ field after its vev
is subtracted. It can then decay into final states with $B=\pm 2$.

On the way to calculating the baryon asymmetry, let us first discuss
the out of equilibrium condition. As the temperature of the universe
falls below the masses of the $X,Y,Z$ particles, the annihilation
processes $X\bar{X}\rightarrow d^c\bar{d^c}$ (and analogous
processes for $Y$ and $Z$) remain in equilibrium. As a result, the
number density of $X,Y,Z$ particles gets depleted and only the $S$
particle survives along with the usual standard model particles. One
of the primary generic decay modes of $S$ is $S\rightarrow
u^cd^cd^cu^cd^cd^c$. There could be other decay modes which depend on
the details of the model. Those can be made negligible by choice of
parameters which do not enter our discussion of $N-\bar{N}$ and
baryogenesis.  For $T\geq M_S$, the decay rate of $S$ is given by 
$\Gamma_S\sim
\frac{T^{13}}{16\pi^9M^{12}_X}$ where we have set the masses of
$X,Y$ and $Z$ particle to be of same order for simplicity. This
decay goes out of equilibrium around $T_*\simeq
M_X\left(\frac{160\pi^9M_X}{M_{P\ell}}\right)^{1/11}$. Here we have
assumed that the coupling of the $X,Y,Z$ particles to second and
third generation quarks are of order 0.1-1. This gives $T_*\sim
0.1-0.2 M_X$ or in the sub-TeV range. Below this temperature the decay 
rate
of $S$ falls very rapidly as the temperature cools. However as soon
as $T\leq M_S$, the decay rate becomes a constant whereas the
expansion rate of the universe is slowing down. So at a temperature
$T_d$, $S$ will start to decay at
\begin{equation}
T_d \simeq (\frac{M_{P\ell}M_S^{13}}{(2\pi)^9M_X^{12}})^{1/2}
\end{equation}
Since the corresponding epoch must be above that of big bang
nucleosynthesis, this puts a constraint on the parameters of the
model. For instance, for $M_S\sim 200 $ GeV and $M_X\sim 3$ TeV, we
get $T_d\sim $ GeV. Also this implies that the $X,Y,Z$ masses cannot be 
arbitrarily high, since the heavier these particles are, the lower $T_d$ 
will be. We expect this upper limit to be in the TeV range at most.

It is well known that baryon asymmetry can arise only via the
interference of a tree diagram with a one loop diagram. The tree diagram
is clearly the one where $S\rightarrow 6 q$. There are however two classes
of loop diagrams that can contribute: one where the loop involves the same
fields $X,Y$ and $Z$. A second one involves W-exchange, which involves
only standard model
physics at this scale (Fig. 3). We find that the second contribution can 
actually
dominate. It also has the advantage that it involves less number of
arbitrary parameters. The
baryon asymmetry is defined as follows
\begin{eqnarray}
\epsilon_B\simeq \frac{n_S}{n_\gamma}\frac{\Gamma(S\rightarrow
6q)-\Gamma(S\rightarrow 6\bar{q})}{\Gamma(S)}
\end{eqnarray}

\begin{figure}[h!]
\includegraphics[scale=0.8]{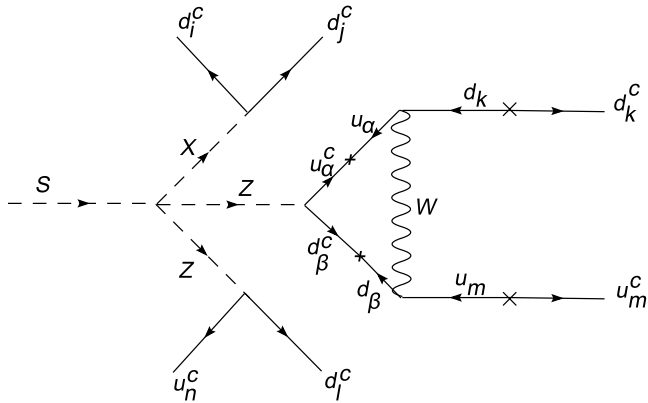}
\caption{One loop diagram for S decays.}
\end{figure}
We find that
\begin{equation}
\epsilon_B \simeq \left\{ \begin{array}{ll}
2\alpha_2Im(V_{tb}V^*_{cb}h_{33}h^*_{23})\frac{m_cm_b^2m_t}{M_W^2M_S^2};
\;\;\;\;\;\; M_S<m_t   \\
2\alpha_2\frac{m_sm_bm_t^2}{M_W^2M_S^2}Im[(h_{33}h^*_{32})(V^*_{tb}V_{cb})];\;
M_S>m_t&
        \end{array} \right.
\label{asym}
\end{equation}
 Note that the trace in the above equation has an imaginary part
 and therefore leads to nonzero asymmetry. The magnitude of the
asymmetry depends on $T_d/M_S$ as well as the detailed
profile of the various coupling matrices $h,g,f$ and we can easily
get the desired value of the baryon asymmetry by appropriately
choosing them. We have checked that there is no conflict between the
desired magnitude of baryon asymmetry and the present lower bound on
the $N-\bar{N}$ transition time of $10^{8}$ sec.

An important point about our baryogenesis mechanism, is that as 
the masses of the
 $X,Y,Z$ particles get larger, the amount of baryon asymmetry goes
 down for given $M_S$ (due to the dilution factor $T_d/M_S$ discussed 
below) as does the strength of the $\Delta B=2$ transition 
  giving
 interesting correlation between the $N-\bar{N}$ process and
 baryon asymmetry. In fact an adequate baryogenesis puts 
an upper limit on $\tau_{N-\bar{N}}$ as discussed below.

 Using $m_c(m_c) = 1.27\;GeV$, $m_b(m_b) = 4.25\;GeV$, $m_t=
 174\;GeV$, $V_{cb} \simeq 0.04$, $M_S=200\;GeV$ and $|h_{33}| \simeq 
|h_{23}|\$
 1$ we find $\epsilon_B \sim 10^{-8}$.

There is a further dilution of the baryon asymmetry arising from the
fact that $T_d \ll M_S$ since the decay of $S$ also releases entropy
into the universe. In this case the baryon asymmetry reads
\begin{equation}
\eta_B \simeq \epsilon_B\frac{T_d}{M_S}
\end{equation}
In order that this dilution effect is not excessive, there must be a
lower limit on the ratio $T_d/M_S$.

From our estimate above we require that $T_d/M_S \geq 0.01$. Since
the decay rate of the $S$ boson depends inversely as a high power of
$M_{X,Y}$, higher $X,Y$ bosons would imply that $\Gamma_S \sim H$ is
satisfied at a lower temperature and hence give a lower $T_d/M_S$.
In figure $3$ we plotted $M_{X,Y}$ vs $M_S $ using $T_d\geq
M_S/100$, and the constraint $G_{N\bar{N}} \leq 10^{-28}\;GeV^{-5}$.
The coupling $\bar{\lambda}^4 \equiv \lambda_1h_{11}^2f_{11} \sim
\lambda_2hg_{11}^2$ . This in turn implies that the
$\tau_{N-\bar{N}}$ must have an upper limit. For instance, for
choice of the coupling parameters $\lambda\sim f\sim h \sim g \sim
10^{-3}$, and $M_S \simeq 200\; GeV$ we find $\tau_{N-\bar{N}}\leq
10^{10}\;sec $.

\section{Conclusion}

In conclusion, we have presented a realistic quark-lepton unified
model where despite the high seesaw ($v_{BL}$) scale (in the range
of $\sim 10^{12}$ GeV), the $N-\bar{N}$ oscillation time can be
around $10^{10}$ sec. due to the presence of a new supersymmetric
graph and accidental symmetries of the Higgs potential (also
connected to supersymmetry). This oscillation time is within the
reach of possible future experiments. We have also found a new way to 
generate the baryon asymmetry of the universe for the case when 
$N-\bar{N}$ oscillation is observable.
These results should provide a motivation to conduct a new round of 
search for $N-\bar{N}$ oscillation.

I would like to thank K. S. Babu, B. Dutta, Y. Mimura and S. Nasri for 
collaborations that led to the results reported in this talk. I like to 
thank Y. Kamyshkov for many discussions on the experimental prospects for 
$N-\bar{N}$ oscillation. I would also like to thank colleagues at the 
Institute of Physics, Bhubaneswar for kind hospitality during the WHEPP9 
symposium. This work of R. N. M. is supported by 
the National Science Foundation grant no. Phy-0354401.


\begin{thebibliography}{99}
\bibitem{thooft}
  G.~'t Hooft,
  Phys.\ Rev.\ Lett.\  {\bf 37}, 8 (1976).


\bibitem{sakharov}
  A.~D.~Sakharov,
  JETP Lett.\  {\bf 5}, 24 (1967). 


\bibitem{ps}
  J.~C.~Pati and A.~Salam,
  Phys.\ Rev.\ D {\bf 10}, 275 (1974).


\bibitem{gg}
H.~Georgi and S.~L.~Glashow,
  Phys.\ Rev.\ Lett.\  {\bf 32}, 438 (1974).


\bibitem{pdecay}
  S.~Dimopoulos, S.~Raby and F.~Wilczek,
  Phys.\ Lett.\ B {\bf 112}, 133 (1982).


\bibitem{kuzmin}
V.~A.~Kuzmin, JETP Lett. {\bf 12}, 228 (1970).


\bibitem{glashow}
S. L. Glashow, Cargese Lectures (1979).


\bibitem{marshak}
  R.~N.~Mohapatra and R.~E.~Marshak,
  Phys.\ Rev.\ Lett.\  {\bf 44}, 1316 (1980).
\bibitem{seesaw}
P. Minkowski, Phys. lett. B {\bf 67}, 421 (1977); M.~Gell-Mann,
P.~Ramond, and R.~Slansky, \emph{Supergravity} (P.~van
  Nieuwenhuizen et al. eds.), North Holland, Amsterdam, 1979, p.~315;
T.~Yanagida, in \emph{Proceedings of the Workshop on the Unified
  Theory and the Baryon Number in the Universe} (O.~Sawada and
  A.~Sugamoto, eds.), KEK, Tsukuba, Japan, 1979, p.~95;
S.~L. Glashow, \emph{The future of elementary particle physics},
in
  \emph{Proceedings of the 1979 Carg{\`e}se Summer Institute on Quarks and
  Leptons} (M.~L{\'e}vy et al. eds.), Plenum Press, New York, 1980, p.~687;
R.~N. Mohapatra and G.~Senjanovi{\'c}, Phys. Rev. Lett. {\bf 44},
912 (1980).

\bibitem{babu}  For a high scale seesaw model where $N-\bar{N}$
oscillation is observable, see K.~S.~Babu and R.~N.~Mohapatra,
  Phys.\ Lett.\ B {\bf 518}, 269 (2001); in this model however,
the lightest SUSY particle, the neutralino is unstable and cannot
be a dark matter candidate. Also leptogenesis cannot explain the
origin of matter, since fast $\Delta B =2$ interactions erase any
baryons generated via this mechanism.

\bibitem{DMM} B.~Dutta, Y.~Mimura and R.~N.~Mohapatra,
  Phys.\ Rev.\ Lett.\  {\bf 96}, 061801 (2006)
  

\bibitem{chacko}
  Z.~Chacko and R.~N.~Mohapatra,
  Phys.\ Rev.\ D {\bf 59}, 055004 (1999)
  [hep-ph/9802388].

  \bibitem{babu1} K. S. Babu, R. N. Mohapatra and S. Nasri, to
  appear; see also talk by K. S. Babu, at this symposium.

\bibitem{fuku}
  M.~Fukugita and T.~Yanagida,
  Phys.\ Lett.\ B {\bf 174}, 45 (1986);
%
  V.~A.~Kuzmin, V.~A.~Rubakov and M.~E.~Shaposhnikov,
  Phys.\ Lett.\ B {\bf 155}, 36 (1985).

\bibitem{milla}
  M.~Baldo-Ceolin {\it et al.},
  Z.\ Phys.\ C {\bf 63}, 409 (1994).


\bibitem{nnbar}
  M.~Takita {\it et al.}  [KAMIOKANDE Collaboration],
  Phys.\ Rev.\ D {\bf 34}, 902 (1986);
%
  J.~Chung {\it et al.},
  Phys.\ Rev.\ D {\bf 66}, 032004 (2002)
  [hep-ex/0205093].

\bibitem{kamyshkov}
  Y.~A.~Kamyshkov,
  hep-ex/0211006.

\bibitem{Rao:1982gt}
  S.~Rao and R.~Shrock,
  Phys.\ Lett.\ B {\bf 116}, 238 (1982);
%
  J.~Pasupathy,
  Phys.\ Lett.\ B {\bf 114}, 172 (1982);
%
  Riazuddin,
  Phys.\ Rev.\ D {\bf 25}, 885 (1982);
S.~P.~Misra and U.~Sarkar,
  Phys.\ Rev.\ D {\bf 28}, 249 (1983).






\end{thebibliography}
\end{document}